# Computational Studies on Structural and Excited State Properties of Modified Chlorophyll *f* with Various Axial Ligands

*S R K Chaitanya Sharma Y[a], Ganga Periyasamy[a, b] and Swapan K Pati* *[b, c]*

(a) Chemistry and Physics of Material Unit, (b) Theoretical Science Unit, (c) New Chemistry Unit Jawaharlal Nehru Centre for Advanced Scientific Research, Jakkur P.O, Bangalore-560 064, India. Fax: 91(80) 2208 2766/2767, E-mail: pati@jncasr.ac.in

**Abstract**

Time Dependent Density Functional Theory (TDDFT) calculations have been used to understand the excited state properties of "modified-chlorophyll" *f* ('Chlide' *f*), Chlide *a*, Chlide *b* and axial ligated (with Imidazole, $H_2O$, $CH_3OH$, $CH_3COOH$, $C_6H_5OH$) Chlide *f* molecules. The computed differences among the $Q_x$, $Q_y$, $B_x$ and $B_y$ band absorbance wavelengths of Chlide *a*, *b* and *f* molecules are found to be comparable with the experimentally observed shifts for these bands in 'chlorophyll' *a* ('chl' *a*), chl *b* and chl *f* molecules. Our computations provide an evidence that the red shift in $Q_y$ band of chl *f* is due to the extended delocaliztion of macrocycle chlorin ring due to the presence of –CHO group. The local contribution from the –CHO substituent to the macrocycle chlorin ring stabilizes the corresponding molecular orbitals (LUMO of the Chlide *f* and LUMO-1 of the Chlide *b*). All the absorption bands of Chlide *f* shift to higher wavelengths on the addition of axial ligands. Computed redox potentials show that, among the axial ligated Chlide *f* molecules, Chlide *f* -Imidazole acts as a good electron donor and Chlide *f* –$CH_3COOH$ acts as a good electron acceptor.

**Introduction**

Chlorophylls (chls) are the essential light harvesting pigments of photo-systems present in the thylakoid membrane of plants, algae and cyanobacteria.[1,2] Five chemically distinct chls are known from oxygenic photosystems, termed as chl *a*, chl *b*, chl *c*, chl *d* and chl *f* (see scheme 1), which absorb solar light at different wavelengths.[3,4] All chls contain a chlorin ring with Mg at its centre (bonded to the nitrogens of the four pyrrole units present in the chlorin ring) and they differ from each other by their substituents R, R′, R″ positions as shown in scheme 1. These substituents affect the π– electron conjugation of the chlorin ring, and thus lead to different absorption spectra.[5]

Electronic spectral properties of chls (except chl *f*) or their models have been studied by several researchers[4,6-20] for many years at different levels of theory, like ZINDO/S-CIS[12], SAC-CI[10], *ab-initio*[13,15], TDDFT[17-19], CASPT2[7]. In these studies, four important peaks were observed in the absorption spectra of chls, which include low energy $Q_x$, $Q_y$ bands and high energy $B_x$, $B_y$ bands. The $Q_y$ absorption wavelengths of the chls are: chl *a* (662 nm), chl *b* (644 nm), chl *d* (688 nm) and chl *f* (706 nm).[3,13,21-23] Among all the chls, the $Q_y$ absorption maximum of chl *f* is red–shifted and it is in the near infrared (NIR)–region. Chen *et. al.,*[21] discovered chl *f* from cyanobacteria and reported preliminary structural and spectroscopic studies. However, the reason behind the huge red–shift in the $Q_y$ band in chl *f*, when compared with other chls has not been understood. In the present study, we focus our attention on microscopic understanding behind the unusual spectroscopic signatures of chl *f*, and to compare and contrast various properties of chl *f* with other chls.

Although all chls are interesting, our study only focuses on chl *f* and its structural isomer chl *b*. Chl *b* and chl *f* structurally differs from each other only by the position of – CHO group (see scheme 1). This makes the system interesting for electronic structural studies, because just by changing the position of a functional group, there is a huge red-shift in the $Q_y$ band (54 nm).[21] For a proper microscopic understanding of the structural and spectroscopic properties of these systems, we have employed the *ab initio* methods such as density functional theory (DFT) and time dependent version of it (TDDFT).

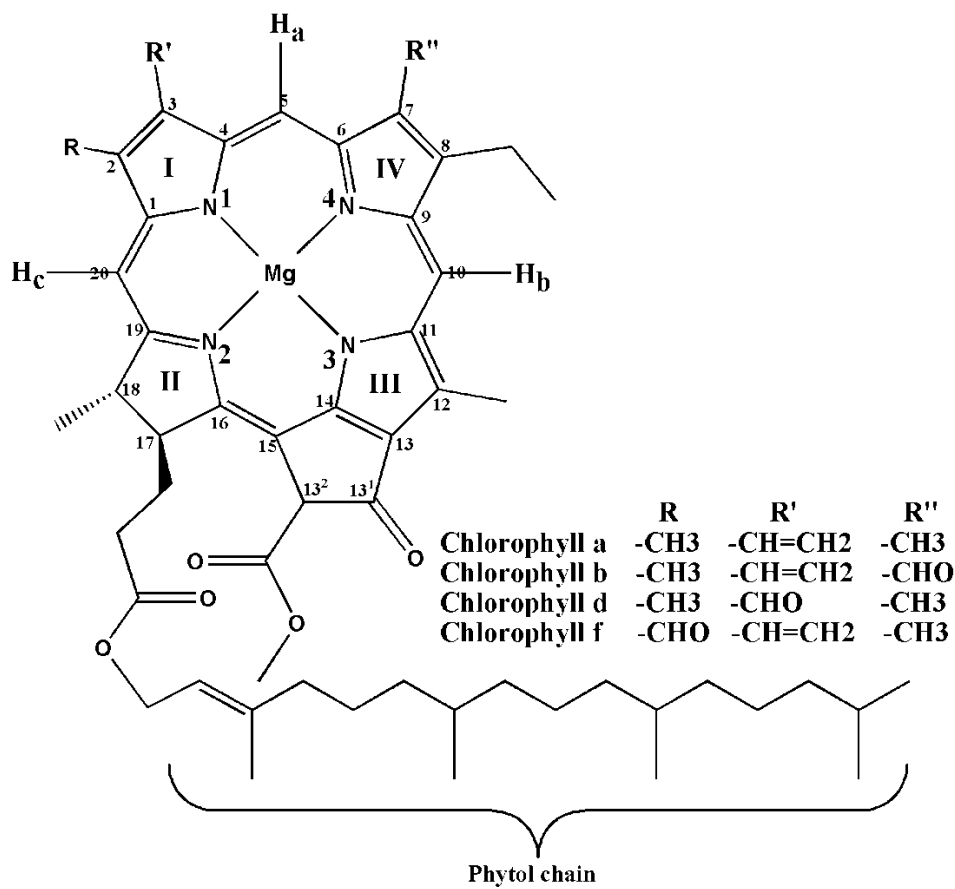

**Scheme 1**. Schematic representations of chemically distinct chlorophylls. Numbering scheme used in this article is shown.

To determine the structure of a chl within the complex protein environment is a vast challenge. In many chl proteins, the Mg–atom exhibits penta-coordination,[24,25] where the fifth coordination may be with an amino acid residue or with a water molecule. There are a few theoretical papers which considered the effects of axial ligation in chl *a*.[6,13-15,26-28] To our knowledge no work has been done on the affect of axial ligation to chl *f*. The present study focuses on the structural and excited state properties of Chlide *f* in the presence and absence of axial ligands (such as Imidazole, $CH_3COOH$, $CH_3OH$, $C_6H_5OH$ and $H_2O$). The obtained structural changes are well supported by the computed NMR results. We also have studied the redox potentials of these molecules, since chls in photosystem are involved in electron transfer.

**Computational Details**

Initial structures of chl *f* and chl *b* are created from the XRD structure of chl *a* found at 2.5 Å resolutions (1JBO.pdb).[29] The substituent at the 17th carbon of the chlorin ring (pyrrole ring II) is modelled as –$CH_3$ group in order to reduce the computational cost (see scheme 1 and figure 1). Removal of phytol chain is considered as a reasonable approximation from several previous computational studies on chlorophyll *a,* which shows negligible changes in the absorption and redox prorperties of the chl molecules in the absence of phytol chain.. [6,13,15,19,20,27,28] We refer this modified chlorophyll structure (a chlorophyll whose substituent at 17th carbon of the chlorin ring is modelled with methyl group) as "Chlide". We also have performed calculations considering complete structure of chlorophylls to know the affects of the substituent at the 17th carbon on the absorption properties of chlorophylls and, as expected, we find little changes in the absorption spectra. (see figures S3, S4 and Table S1 for comparison).

As there are two conformers possible with respect to the –CHO group for Chlide *f* and Chlide *b*, geometry optimizations are performed to find the conformer with minimum energy and this minimum energy conformer is considered for further calculations. Five neutral axial ligands are considered in our study. 'Histidine' is modelled as "Imidazole", 'Aspartate' as "$CH_3COOH$", 'Serine' as "$CH_3OH$", 'Tyrosine as "Phenol" and 'water' as "$H_2O$". Considering functional groups instead of complete amino acids have been proved to mimic the role of amino acids in various axial ligated chl *a* molecules. [27,28] The above mentioned axial ligands occur frequently in photosystems of cyanobacterium.[27]

All the geometry optimizations and energy calculations are performed with the Density Functional Theory (DFT). The three–parameter hybrid functional B3LYP (Beke exchange with Lee, Yang and Parr correlation) [30-32] with 6–31+G (d) basis set is used for all the atoms as implemented in the Gaussian 03 program package. [33] We choose the B3LYP functional, because, it is found to be one of the appropriate functional[14,15,17,19,20,26,27,34] for the prediction of electronic structure for the various oxidation states of chls. We are aware that this DFT functional is not fully adequate to describe the long–range dispersion forces or stacking interactions, but, we do not expect them to play an important role in the single unit Chlide molecules investigated here. The computed energies are corrected for basis set superposition error (BSSE) using the counterpoise method.[35,36] All the molecular properties are calculated at the same level of theory and using the Gaussian 03 program package.[33] $^1$H–NMR chemical shielding values are calculated using Gauge Including Atomic Orbital (GIAO) method.[37-41] Time Dependent DFT (TDDFT)[42,43] is used to calculate the absorption properties of the optimized geometries.

In this work we did not include the implicit solvent effect, since it is already proven that in the continuum model, solvent effect on the neutral ligated chl *a* molecules are negligible[27] and the initial molecular structures are obtained from 1JBO.pdb crystal structure, where all environmental effects have implicitly been included. In addition, it is known that in the protein environment, the inter–molecular H–bonding interaction between axial ligands and the surrounding environment will be dominant, which cannot be mimicked using implicit solvent model. To use an explicit solvent will be expensive and hence, in this work we report only the gas phase results.

**Results and Discussions**

In what follows, initial structural differences in Chlide *f*, Chlide *b* and Chlide *a* are reported along with the structural changes upon axial ligation, in section I. Bond dissociation energies of axial ligated Chlide *f* molecules are reported in section II. Redox properties are discussed in section III. Section IV is devoted to the changes in the absorption spectra among different Chlide molecules, in particular, the reasons for the red–shift in Chlide *f*.

**I. Structural changes**

In this section, we discuss the structural differences among Chlide *f*, Chlide *a* and Chlide *b* along with a structural comparison between the two conformers of Chlide *f*. In addition, structural changes due to axial ligation are also discussed.

Among the many conformers of chl *b*, the two conformers, S–cis chl *b* (chl *b*′) and S–trans–chl *b* (chl *b*) are reported to be more stable with close in energy.[20] We also found the same, where the S–trans conformer turns out to be more stable than S–cis conformer, however, only by 5 kcal mol$^{-1}$, and hence, both conformers can be present in the gas phase. However, one of the conformer might be stabilized in the protein environment due to the presence of an inter–molecular hydrogen bonding interaction. In order to find the conformer present in the nature, the chemical shift values of 'formyl hydrogen' for both conformers are computed. Chemical shift value of S–trans Chlide *b* (11.35 *ppm*) is comparable with experimental value of chl *b* (11.25 *ppm*) (see Table 1), than the S–cis Chlide *b* (11.85 *ppm*). In addition, our computed values are also in good agreement with the already reported values using CAM–B3LYP functional [21] which validates the functional and basis set used in this study.

Similarly, S–trans conformer of Chlide *f* is found to be stable by 1 kcal mol$^{-1}$ than the S–cis conformer and the chemical shift value of formyl group for S–cis (11.71 *ppm*) and S–trans (11.51 *ppm*) conformers of Chlide *f* (see Table 1) shows that the S–trans conformer chemical shift value is more comparable with experimental value, which is 11.35 *ppm*. [21] The above studies on Chlides provide us an evidence that chl *b* and chl *f* might be present in the S–trans conformation in protein environment, and are considered for further studies reported in this article.

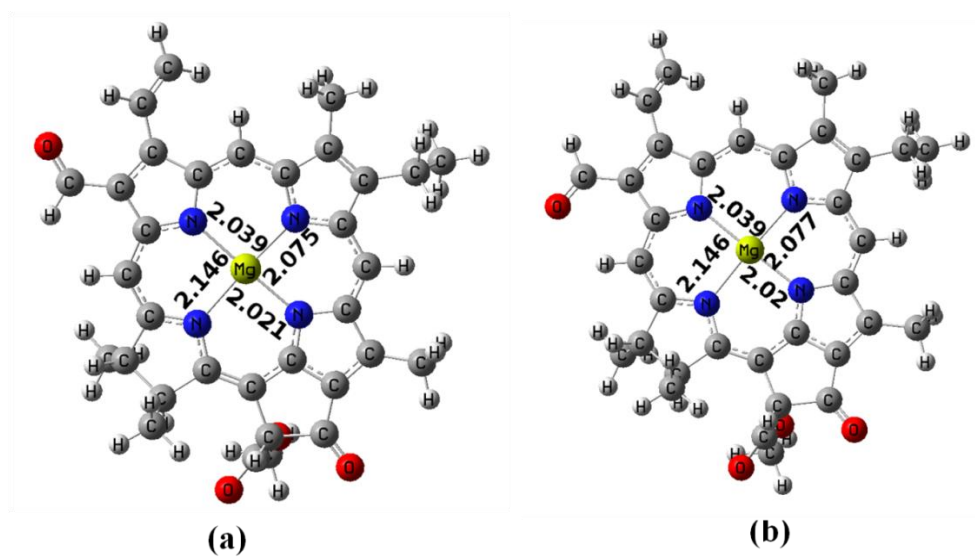

**Figure 1.** Optimized geometry of two stable conformers of Chlide *f* (a) S–cis Chlide *f* (b) S–trans Chlide *f*. Important bond distances are shown in Å. Note that the Mg is present in the plane of macrocycle.

| Molecule | Method | H$_a$ | H$_b$ | H$_c$ | Formyl hydrogen |
|---|---|---|---|---|---|
| Chlide *a* | DFT | 9.28 | 9.67 | 8.51 | — |
| Chl *a* | Exp[44] | 9.29 | 9.54 | 8.32 | — |
| Chlide *b* | DFT | 10.93 | 9.48 | 8.17 | 11.35 |
| Chl *b* | Exp[45] | 10.04 | 9.64 | 8.20 | 11.22 |
| Chlide *f* | DFT | 9.60 | 9.57 | 10.17 | 11.51 |
| Chl *f* | Exp[21] | 9.79 | 9.86 | 9.77 | 11.35 |
| Chlide *b′* | DFT | 9.73 | 9.71 | 8.31 | 11.83 |
| Chlide *f′* | DFT | 9.84 | 9.64 | 8.95 | 11.71 |
| $H_2O$ | DFT | 9.57 | 9.54 | 10.20 | 11.55 |
| Imidazole | DFT | 9.60 | 9.48 | 10.15 | 11.49 |
| $CH_3OH$ | DFT | 9.59 | 9.63 | 10.09 | 11.45 |
| $CH_3COOH$ | DFT | 9.44 | 9.66 | 10.21 | 11.16 |
| $C_6H_5OH$ | DFT | 9.43 | 9.77 | 10.15 | 11.17 |

**Table 1** Computed $^1$H NMR chemical shift values (*ppm*) for the $H_a$, $H_b$, $H_c$ and formyl hydrogen atoms of various Chlides studied in this paper and the corresponding shifts in the experimentally studied chls. Hydrogen labelling schemes are shown in the scheme 1. Tetramethylsilane (TMS) is used as a reference.

In the optimized structure of Chlide *f*, Mg is in the plane of the chlorin macrocycle, and is coordinated asymmetrically to the four pyrrole nitrogens. The order of the Mg–N distances are Mg–N2 (2.15 Å) > Mg–N4 (2.08 Å) > Mg–N1 (2. 04 Å) > Mg–N3 (2.02 Å), (see Table 2) which is similar to the order found for other chls obtained using B3LYP[6,12,14,15,20,26-28] and also comparable with the experimental Mg-N bond distance order obtained for ethyl chlorophyllidine *a* (where the differences in the experimental and calulated bond distances are within 0.02 Å).[46] Longer distance of Mg–N2 is because of the saturation of the corresponding pyrole ring (ring II).

The affect of the functional groups (R, R′ and R″) on chlorin ring can further be understood by comparing the structure of Chlide *f* with those of Chlide *a* and Chlide *b* (see scheme 1). The affect of the changes in the functional groups are clearly reflected in the Mg– N1 and Mg–N4 bond distances and $^1$H–NMR chemical shift values of $H_a$ and $H_c$ as shown in the tables 1 and 2. The Mg–N1 bond in Chlide *f* (2.04 Å) is longer than the Mg–N1 bond in Chlide *a* (2.03 Å) and Chlide *b* (2.03 Å), due to the presence of the electron withdrawing –CHO substituent at the corresponding pyrole ring (ring I) of Chlide *f* (see Table 2). Presence of the – CHO substituent at ring I in Chlide *f* is also the reason for the downfield shift of $H_c$ by 1.66 (2.00) *ppm* when compared to Chlide *a* (Chlide *b*). Similarly, the presence of –CHO group at ring IV in Chlide *b* results in longer Mg–N4 bond (2.08 Å) than Chlide *f* (2.08 Å) and Chlide *a* (2.07 Å) (see Table 2). In addition, $^1$H–NMR chemical shielding value of the $H_a$ of Chlide *b* is shifted downfield by 1.33 (1.65) *ppm* when compared to the $H_a$ hydrogen of the Chlide *f* (Chlide *a*). (see Table 1)

These results provide an evidence for the most possible chl *f* conformer in the gas phase. However, the Mg atom of chl *f* could be penta–coordinated in protein environment, as observed and studied earlier in chl *a* and chl *b*. [6,12,14,15,26-28,47] As mentioned in the computational section, we have considered five neutral ligands for our studies.

Coordination of Mg with neutral ligands does not change the order of Mg–N bond distances: Mg–N2 > Mg–N4 > Mg–N1 > Mg–N3 (see Table 2), however, Mg–N bonds are elongated by 0.01–0.04 Å and Mg atom is displaced from the plane of the macrocycle by 0.2 – 0.4 Å (see figure 2). The amount of the displacement of Mg atom is shown by computing out of plane distance ($d_{OOP}$) (see Table 2) by,

$$d_{OOP} = \frac{\left[\cos(\theta/2) * (d_1 + d_3)\right]}{2}$$

where, $\theta$ is the N1–Mg–N3 bond angle; $d_1$ = Mg–N1 bond distance; and $d_2$ = Mg–N3 bond distance. $d_{OOP}$ distances of Mg are in the order: Mg–Imidazole [Mg–N] (0.37 Å) > Mg–CH$_3$OH [Mg–O] (0.27 Å) > Mg–C$_6$H$_5$OH [Mg–O] (0.26 Å) ≈ Mg–CH$_3$COOH [Mg–O] (0.25 Å) > Mg–H$_2$O [Mg–O] (0.24 Å). The obtained $d_{OOP}$ of Mg in Mg-H$_2$O is comparable with the experimentally obtained $d_{OOP}$ (0.385 Å) of Mg in ethyl chlorophyllide *a* dihydrate. [46] $d_{OOP}$ distance highly depends on the steric class of interactions of the ligated groups (see Table 2). There is one exception from the trend, CH$_3$OH, for which $d_{OOP}$ distance is larger than $C_6H_5OH$ and $CH_3COOH$. This is due to orientation of $CH_3$ group of $CH_3OH$, which is very close to the macrocycle (see figure 2d).

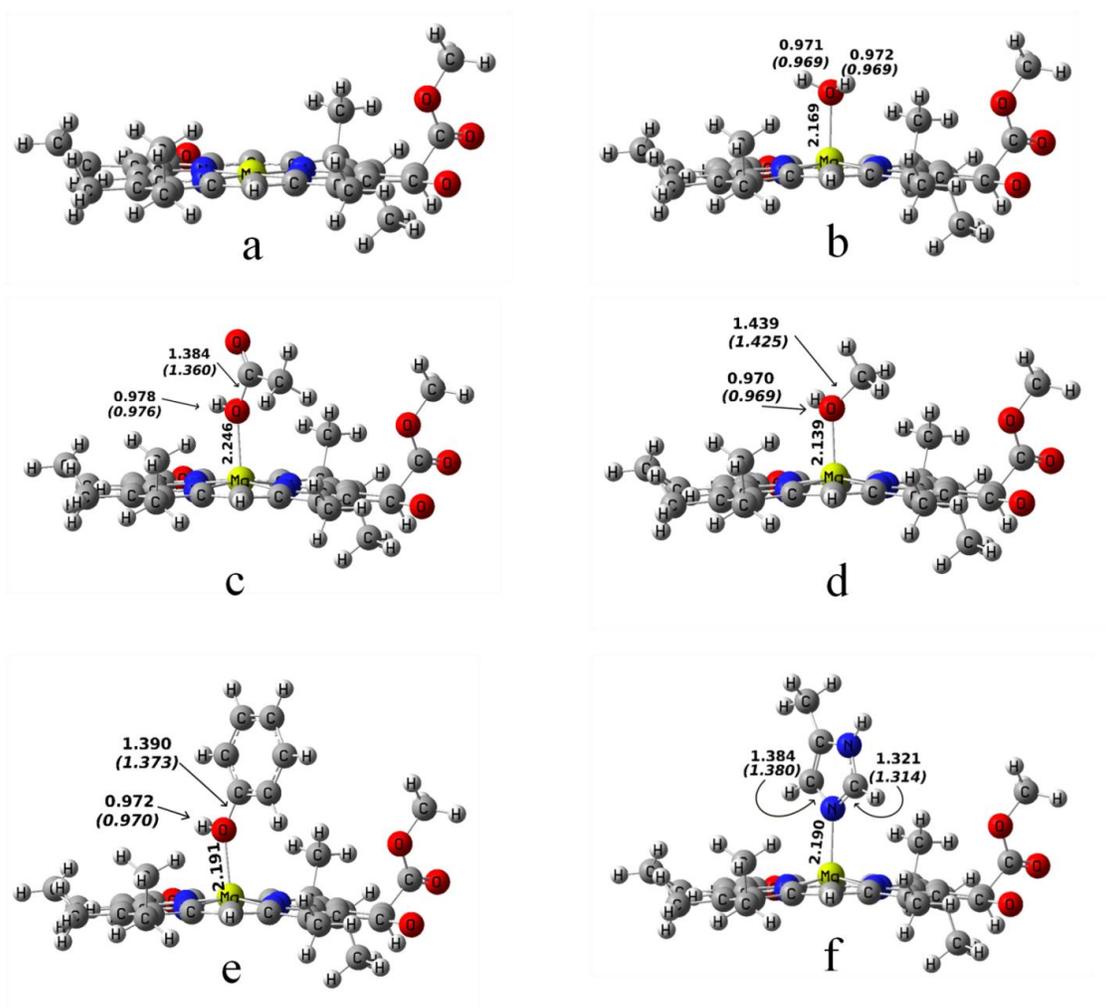

**Figure 2.** Optimized geometry of (a) Chlide *f*, axial ligated, (b) $H_2O$, (c) $CH_3COOH$, (d) $CH_3OH$, (e) $C_6H_5OH$ and (f) Imidazole Chlide *f* molecules. Important bond distances are shown in Å. Distances of the respective bonds in the bare ligands are written in italics within brackets. The displacement of Mg from the plane can be seen for axial ligated Chlide *f* molecules and the corresponding distances ($d_{OOP}$) are shown in the Table 2.

| Molecules | Mg–N1 | Mg–N2 | Mg–N3 | Mg–N4 | Mg–ligand | $d_{OOP}$ | Mulliken partial atomic charge (in e) | |
|---|---|---|---|---|---|---|---|---|
| | | | | | | | Mg | Ligated atom |

| Ligand | | | | | | | | |
|---|---|---|---|---|---|---|---|---|
| Chlide *a* | 2.03 | 2.15 | 2.02 | 2.07 | – | 0.02 | 0.84 | – |
| Chlide *b* | 2.03 | 2.15 | 2.02 | 2.08 | – | 0.03 | 0.86 | – |
| Chlide *b′* | 2.03 | 2.15 | 2.02 | 2.08 | – | 0.02 | 0.82 | – |
| Chlide *f* | 2.04 | 2.15 | 2.02 | 2.08 | – | 0.02 | 0.82 | – |
| Chlide *f′* | 2.04 | 2.15 | 2.02 | 2.08 | – | 0.02 | 0.80 | – |
| $H_2O$ | 2.06 | 2.17 | 2.04 | 2.11 | 2.17 | 0.24 | 0.42 | –0.84 (*-0.93*) |
| Imidazole | 2.08 | 2.20 | 2.06 | 2.12 | 2.19 | 0.37 | –2.04 | –0.27 (*-0.31*) |
| $CH_3OH$ | 2.07 | 2.18 | 2.04 | 2.10 | 2.14 | 0.27 | –0.51 | –0.47 (*-0.65*) |
| $CH_3COOH$ | 2.06 | 2.17 | 2.03 | 2.09 | 2.25 | 0.22 | –1.23 | –0.43 (*-0.57*) |
| $C_6H_5OH$ | 2.07 | 2.17 | 2.04 | 2.10 | 2.19 | 0.26 | –1.39 | –0.47 (*-0.69*) |
| Negative ligands | | | | | | | | |
| $CH_3O^-$ | 2.11 | 2.24 | 2.08 | 2.14 | 1.95 | 0.54 | –0.24 | –0.08 (*-0.82*) |
| $CH_3COO^-$ | 2.12 | 2.33 | 2.08 | 2.16 | 1.96 | 0.62 | –0.27 | –0.22 (*-0.63*) |
| $C_6H_5O^-$ | 2.09 | 2.21 | 2.07 | 2.13 | 2.02 | 0.48 | –1.44 | –0.08 (*-0.73*) |

**Table 2.** Computed important bond distances (Å) and Mulliken partial atomic charges (e) for Chlide *a*, Chlide *b*, Chlide *b′*, Chlide *f*, Chlide *f′*, neutral ligand ligated Chlide *f* and negative charge ligand ligated Chlide *f*. Mulliken partial atomic charge on ligated atom in bare ligand are given in italics within brackets.

The computed Mulliken partial atomic charges on Mg atom (of the Chlides) and the ligating atom (of the ligands) clearly shows a ligand to metal charge transfer. Compared with Chlide *f* Mg atom (0.82 e), the axial ligated Chlide *f* Mg gains a negative charge of -0.40e, -2.86e, -1.33e, -2.05e and -2.21 e from $H_2O$, Imidazole, $CH_3OH$, $CH_3COOH$ and $C_6H_5OH$, respectively. Among the five neutral ligands, Imidazole donates more charge to the metal atom because of the direct coordination of the aromatic ring to the Mg-atom of chlorophyll.

Mg–axial ligand distances are in the order: Mg–$CH_3COOH$ [Mg–O] (2.244 Å) > Mg–Imidazole [Mg–N] (2.190 Å) ≈ Mg–$C_6H_5OH$ [Mg–O] (2.190 Å) > Mg–$H_2O$ [Mg–O] (2.168 Å) > Mg–$CH_3OH$ [Mg–O] (2.139 Å). The experimental Mg-$H_2O$ distance in ethyl chlorophyllide *a* dihydrate [46] is 2.035 Å, which is quite comparable with our calculated Mg-$H_2O$ bond distance of 2.168 Å. The ligated oxygen atoms of $CH_3COOH$, $C_6H_5OH$ molecules are attached to the acidic hydrogen, which increases Mg–O bond distances compared with Mg–$OH_2$ distance. However, the larger electrostatic repulsion between the axial -$OCH_3$ hydrogen and chlorin macrocycle decreases the Mg–O bond distance in Chlide *f*–$CH_3OH$ (see figure 2) by distorting Mg atom from its plane. The effect of the presence of acidic hydrogen can further be explained by calculating the Mg–O bond distances for negatively charged $C_6H_5O^-$, $CH_3COO^-$, $CH_3O^-$ ligands. The Mg–O bond distances are in the order: Mg–$C_6H_5O^-$ (2.02 Å) > Mg–$CH_3COO^-$ (1.96 Å) > Mg–$CH_3O^-$ (1.95 Å) (see Table 2), which are smaller than the Mg–$OH_2$ (2.17 Å) bond distance.

Computed Mg–N, $d_{OOP}$, Mg–Ligand bond distances of axial ligated Chlide *f* follow the same trend as chl *a* molecules computed at the B3LYP level,[27] except that these bond distances are shorter in Chlide *f* by 0.01–0.03 Å. These results show that, axial ligation to Chlide *f* affects the macrocycle unit similar to that in chl *a*.

Additionally, axial ligation to Chlide *f* follow the similar trend of the chemical shielding values as in Chlide *f*, *i.e.*, the formyl hydrogen is always in downfield when compared to the $H_a$, $H_b$ and $H_c$. But, there are small changes in the chemical shielding values depending upon the orientation of the axial ligands. The presence of electron donating neutral axial ligands shift the chemical shielding values of $H_a$, $H_c$, formyl hydrogens by 0.02–0.35 *ppm* towards upfield (see Table 1) and $H_b$ by 0.03–0.20 *ppm* towards downfield. There are three exceptions from this trend (see table 1).

1. The inter–molecular hydrogen bonding interaction between axial $H_2O$ and substituent formyl group is the reason for the downfield shift in formyl, $H_c$ and the upfield shift in $H_b$ proton.
2. The orientation of Imidazole shifts the $H_b$ chemical shielding value towards upfield and
3. The H-bonding interaction between the acidic hydrogen of $CH_3COOH$ and –CHO is the reason for the downfield shift in the $H_c$.

## II. Bond Dissociation Energy ($E_{BDE}$) of axial ligated Chlide $f$

Mg–ligand bond dissociation energy is a measure of its bond strength, which is calculated as,

$$E_{BDE} = E_{[Chlide\ f-Ligand]} - E_{[Chlide\ f]} - E_{[Ligand]},$$

where, $E_{[Chlide\ f-Ligand]}$, $E_{[Chlide\ f]}$ and $E_{[Ligand]}$ are the BSSE corrected energies of the corresponding molecules. These are tabulated in table 3.

Negative $E_{BDE}$ values of axial ligated Chlide $f$ molecules, shows their stability. The stability order of ligands (containing oxygen as ligating atom) bonding to Chlide $f$ is: $CH_3OH > H_2O > C_6H_5OH > CH_3COOH$ (see table 3). $E_{BDE}$ values reflect the Mg-ligand bond distance (see table 2), the shorter the bond, larger is the $E_{BDE}$. The $E_{BDE}$ values are in the same order as the Mg–ligand bond distances (see Table 2), as expected.

Imidazole (Histidine model) has large $E_{BDE}$ value in spite of its longer bond distance. The similar larger $E_{BDE}$ value was observed earlier also for imidazole ligated chl $a$.[27] Interestingly, the reason for strong binding energy of imidazole is traced back due to the direct coordination between aromatic ring and Mg. This direct coordination between imidazole and Mg lead to the large charge transfer (-2.86e) from imidazole to Mg, which is reflected in the Mulliken partial atomic charge (Table 2) at the Mg centre.

| Molecules | RP1 = E (N - 1) –E (N) - 4.43 (eV) | RP2 = E (N) – E (N + 1) - 4.43 (eV) | BSSE corrected $E_{BDE}$ (KJ mol$^{-1}$) |
|---|---|---|---|
| Chlide $f$ | 2.073 | –2.199 | 0.000 |
| $H_2O$ | 1.910 | –2.253 | –45.263 |
| Imidazole | 1.665 | –2.471 | –70.704 |
| CH$_3$OH | 1.856 | −2.308 | −48.624 |
| CH$_3$COOH | 2.019 | −2.144 | −25.654 |
| C$_6$H$_5$OH | 1.883 | −2.253 | −34.184 |

**Table 3.** Computed reduction potential (RP1, RP2 in eV) and BSSE corrected Mg–ligand bond dissociation energies (E$_{BDE}$, KJ mol$^{-1}$) for Chlide $f$ and axial ligated Chlide $f$.

### III. Redox properties of Chlide $f$ and axial ligated Chlide $f$ molecules

One of the major roles of chls in photosystem is that they are involved in electron transfer process, which can be defined by computing reduction potentials (RPs). The reduction potential (RP) is a measure of the ability of a compound to acquire electrons and get reduced. Herein, we have considered a redox reaction with three oxidation states, since a chl can accept/donate electron during the electron transfer process in photosystem

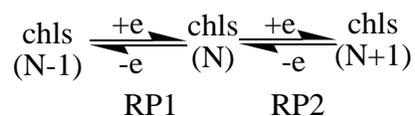

where, N is the number of electrons. RP1 and RP2 are reduction potential for chls (N-1) / chls (N) and chls (N) / chls (N+1) redox pairs respectively. RP1 and RP2 values are computed for relaxed geometries of Chlides using the equations

$$RP1 = E(N-1) - E(N) - 4.43$$

$$RP2 = E(N) - E(N+1) - 4.43$$

where, the factor 4.43 eV is an estimate of the reduction potential of the standard hydrogen electrode [48] and E is the BSSE corrected energy of the corresponding oxidation states. The values are tabulated in table 3.

In general, all the highest occupied and lowest unoccupied molecular orbitals of the Chlide $f$ and axial ligated Chlide $f$ have localized π molecular orbital, (see figure 5 and figure S1), which shows that the redox process has major contributions from the macrocycle ring rather than from the axial ligands. However, addition of axial ligands to Chlide $f$ decreases the RP1 and RP2 values by ~0.05 – 0.4 eV.

Thus, electron donating axial ligands brings stabilization to oxidised state (see Table 3) except $CH_3COOH$, which stabilizes the corresponding reduced (N+1) state (see RP2 value of $CH_3COOH$).

If one compares two compounds, the compound with more positive RP1 or RP2 will acquire the electrons from the other compound and thus acting as an oxidizing agent to the other. Our results provide us an evidence that the $CH_3COOH$ with more positive RP1 and RP2 values can act as electron acceptor and imidazole with less positive RP1 and RP2 values can act as electron donor, when compared with other ligands. In addition, it is clear that the order of RPs is in exactly reverse to the order of $E_{BDES}$ in terms of magnitude. The order of magnitudes of RPs are: $CH_3COOH$ > $H_2O$ > $C_6H_5OH$ > $CH_3OH$ > Imidazole.

**IV. Absorption properties**

There exist large number of computational studies on chl *a*, chl *b* and the related model structures to understand their absorption properties.[8,9,14,49] The effects of the substituents (on the chlorin macrocycle) on the absorption properties of the chls have also been performed.[12-14] TDDFT computed absorption spectra of Chlide *a,* Chlide *b* and Chlide *f* are shown in figure 3a, 3b. The main peaks of the spectrum are the low energy $Q_x$ and $Q_y$ band (600–700 nm), the high energy $B_x$ and $B_y$ bands (350–500 nm). In addition to these bands, we also find a few higher energy bands near the soret region. However, the main emphasis of this work is to understand the underlying reason for the red–shift of experimentally observed $Q_y$ band of chl *f* compared to chl *a* and chl *b*. This can be explained by computing molecular orbital (MO) energies and the MO contributions to each excitation as shown in the Table 4 and figures 5 and 6. Our computed results for $Q_y$ bands in Chlide *a* (589 nm), Chlide *b* (577 nm) and Chlide *f* (615 nm) are lower by ~ 79-91 nm than the experimental transition energies of chl *a* (657 nm), chl *b* (662 nm) and chl *f* (706 nm) due to the absence of protein environment. However, the transition energies follow the same trend as in the experiment *i.e.* Chlide *b* < Chlide *a* < Chlide *f*. Also, the shift in the $Q_y$ bands of Chlide *b* and Chlide *f* with respect to the $Q_y$ bands of Chlide *a* is −12 nm and +26 nm, which are comparable with the experimental shifts of −13 nm and +41 nm, respectively.[3,21] The small differences might be due to the absence of protein environment in our studies. Inclusion of the substituent at C17 to Chlide *f* is found to change the absorption peak values by only 1-2 nm. This can be understood from the molecular orbital plots which show almost no contribution (not shown here) from the phytol chain to the low energy molecular orbitals (see Figure S3, S4 and Table S1)

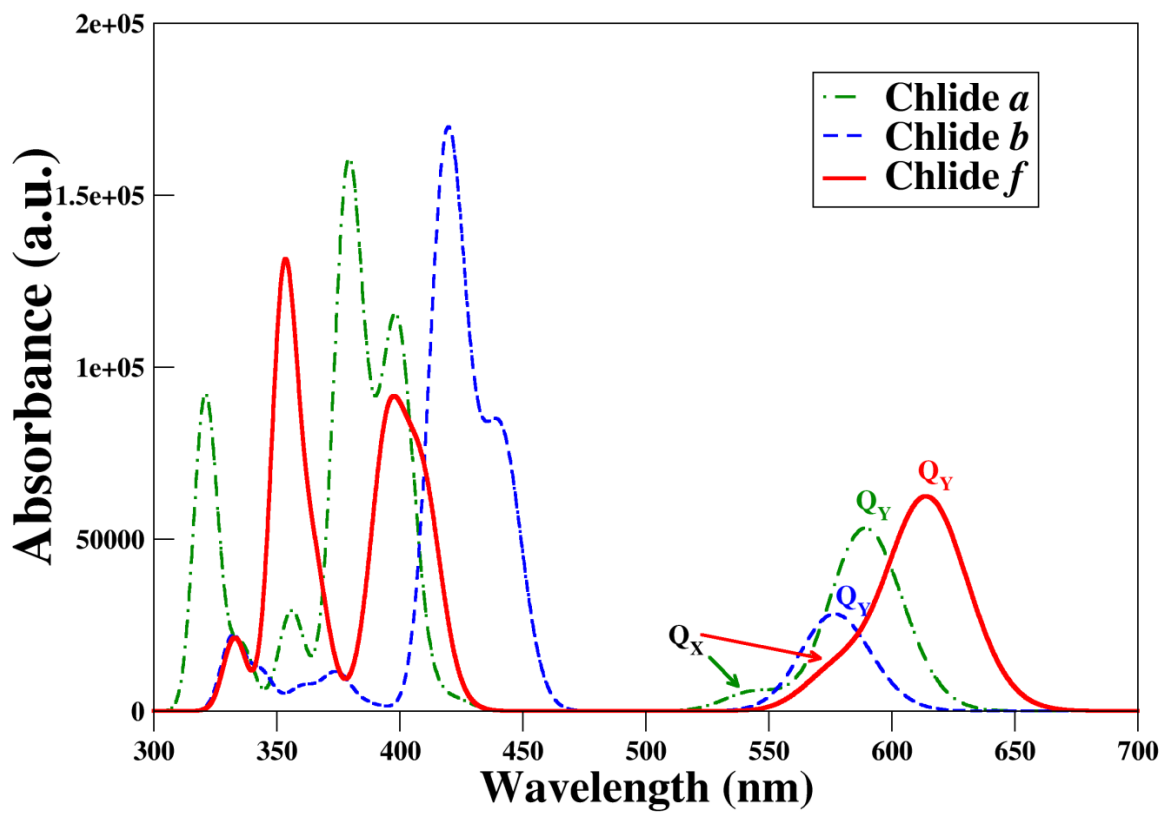

**Figure 3a.** Computed absorption spectra of Chlides *a*, *b* and *f*, where Full Width Half Maximum, FWHM= 1000 cm$^{-1}$.

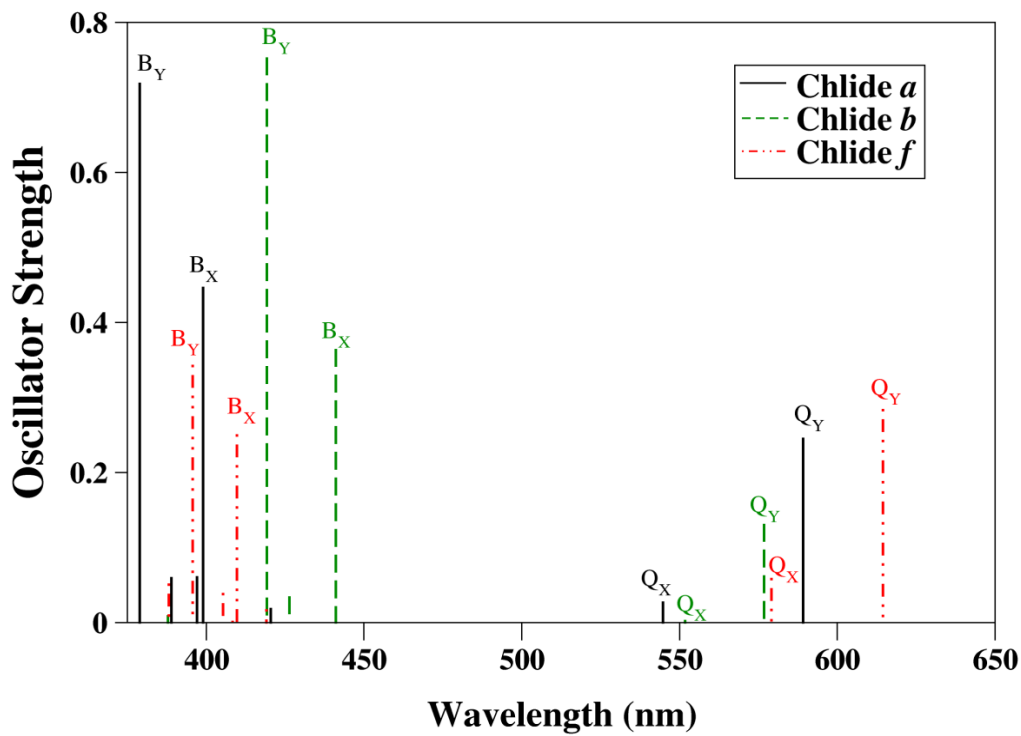

**Figure 3b.** Computed absorption spectra of Chlides *a*, *b* and *f*. Transitions $Q_y$, $Q_x$, $B_x$, $B_y$ are assigned.

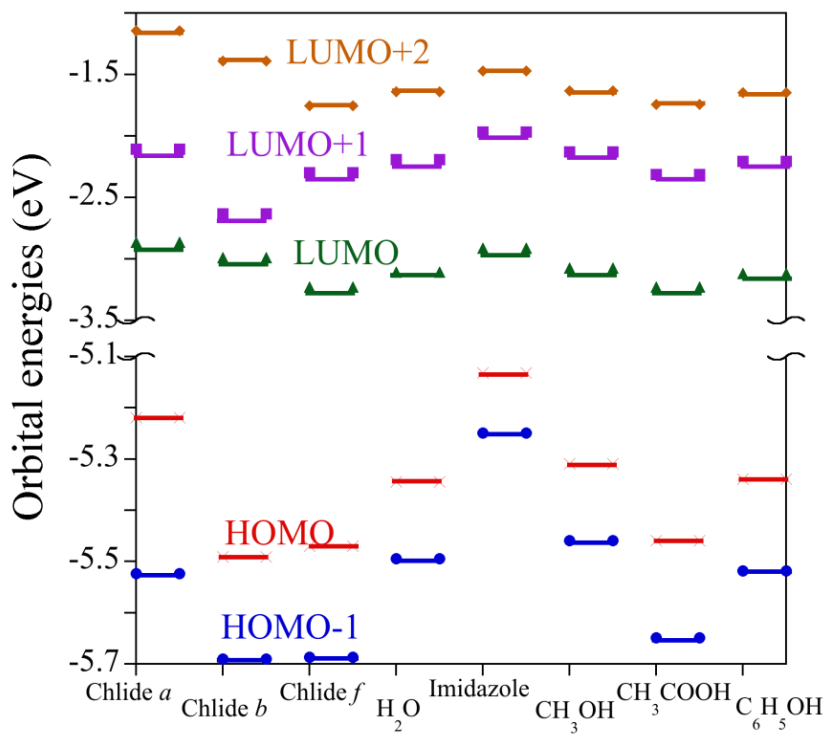

**Figure 4.** HOMO–LUMO energy plots of Chlide *a*, Chlide *b*, Chlide *f*, Chlide *f*-$H_2O$ ($H_2O$), Chlide *f*-Imidazole (Imidazole), Chlide *f*-$CH_3OH$ ($CH_3OH$), Chlide *f*-$CH_3COOH$ ($CH_3COOH$) and Chlide *f*-$C_6H_5OH$ ($C_6H_5OH$) in eV.

| Molecules | Nature | Excitation Wavelength (nm) | Major molecular orbital contributions |
|---|---|---|---|
| Chlide *a* | $Q_y$ | 589 | H to L (0.71) |
| | $Q_x$ | 545 | H-1 to L (0.65) and H to L + 1 (0.31) |
| | $B_y$ | 379 | H–1 to L+1 (0.57) |
| | $B_x$ | 399 | H to L+1 (0.39) and H-3 to L (0.13) |
| Chlide *b* | $Q_y$ | 577 | H–1 to L+1 (0.27) and H to L (0.65) |
| | $Q_x$ | 552 | H-1 to L (0.56) and H to L + 1 (0.40) |
| | $B_y$ | 419 | H–1 to L+1 (0.20) and H to L+1 (0.24) |
| | $B_x$ | 441 | H–1 to L+1 (0.36) and H to L+1 (0.16) |
| Chlide *f* | $Q_y$ | 615 | H to L (0.72) |
| | $Q_x$ | 579 | H-1 to L (0.73) and H to L + 1 (0.19) |
| | $B_y$ | 396 | H–1 to L+1 (0.61) |
| | $B_x$ | 409 | H to L+1 (0.51) |
| $H_2O$ | $Q_y$ | 618 | H to L (0.71) |
| | $Q_x$ | 596 | H-1 to L (0.74) and H to L + 1 (0.16) |
| | $B_y$ | 408 | H–1 to L+1 (0.53) |
| | $B_x$ | 416 | H-2 to L (0.31) and H to L+1 (0.33) |
| Imidazole | $Q_y$ | 622 | H to L (0.71) |
| | $Q_x$ | 609 | H-1 to L (0.74) and H to L + 1 (0.13) |
| | $B_y$ | 413 | H–1 to L+1 (0.42) and H to L+1 (0.24) |
| | $B_x$ | 421 | H to L+1 (0.38) and H-1 to L+1 (0.22) |
| $CH_3OH$ | $Q_y$ | 617 | H to L (0.72) |
| | $Q_x$ | 596 | H-1 to L (0.74) and H to L + 1 (0.15) |
| | $B_y$ | 406 | H–1 to L+1 (0.50) |
| | $B_x$ | 414 | H–1 to L+2 (0.15) and H to L+1 (0.52) |
| $CH_3COOH$ | $Q_y$ | 619 | H to L (0.71) |
| | $Q_x$ | 588 | H-1 to L (0.74) and H to L + 1 (0.13) |
| | $B_y$ | 407 | H–1 to L+1 (0.33) and H-4 to L (0.39) |
| | $B_x$ | 418 | H–3 to L (0.28) and H to L+1 (0.23) |
| $C_6H_5OH$ | $Q_y$ | 620 | H to L (0.72) |
| | $Q_x$ | 593 | H-1 to L (0.73) and H to L + 1 (0.17) |
| | $B_y$ | 403 | H–1 to L+1 (0.32) and H-4 to L (0.39) |
| | $B_x$ | 359 | H–3 to L (0.17) and H to L+1 (0.44) |

**Table 4.** Computed excitation wavelength (nm) for the Chlide *a*, Chlide *b*, Chlide *f* and axial ligated Chlide *f* molecules. Major orbital contributions for the corresponding excitations are also given. In this table 'H' denotes HOMO and 'L' denotes LUMO.

From the TDDFT calculations one can see that the major contributions to the $Q_y$ band involves the excitations from HOMO to LUMO (coefficient value ~ 0.65) and HOMO−1 to LUMO+1 (coefficient value ~ 0.2). The orbital energy differences between the corresponding orbitals on Chlide *a*, Chlide *b* and Chlide *f* clearly reflects these shifts (see figure 4). In the case of Chlide *a*, it is clear that the larger excitation energy is due to the absence of electron withdrawing –CHO group. The presence of electron withdrawing –CHO group stabilizes the orbital energy of LUMO for Chlide *f* and LUMO+1 for Chlide *b* (see figure 5). Among these two, the HOMO–LUMO excitation has major contribution (coefficient value ~ 0.7) in the $Q_y$ band rather than the HOMO−1 to LUMO+1 (coefficient value ~ 0.2). This shows that the specific coordination between the π orbital of macrocycle ring and the –CHO unit is the primary reason for the red shift in Chlide *f* when compared with Chlide *b*.

HOMO and HOMO−1 of both Chlide *f* and Chlide *b* have the same amount of contribution from the − CHO group 15–16 % and 3–4 % respectively, and also, they have similar energy values. However, the LUMO of Chlide *f* has 31 % contribution from − CHO group, which stabilizes the LUMO level by 0.3 eV than its corresponding Chlide *b* LUMO orbital which has only 3 % − CHO group contribution (see figure 5). Similarly, LUMO+1 of Chlide *f* with smaller –CHO contribution (13 %) destabilizes the orbital energy by 0.1 eV compared to the LUMO+1 of Chlide *b*, which has larger (46 %) -CHO contribution. This shows that the MO with more (less) contribution from − CHO group stabilizes (destabilizes) the corresponding level. The LUMO orbital which has larger –CHO contribution in the Chlide *f* has major contribution to the $Q_y$ band which is the main reason for huge red shift.

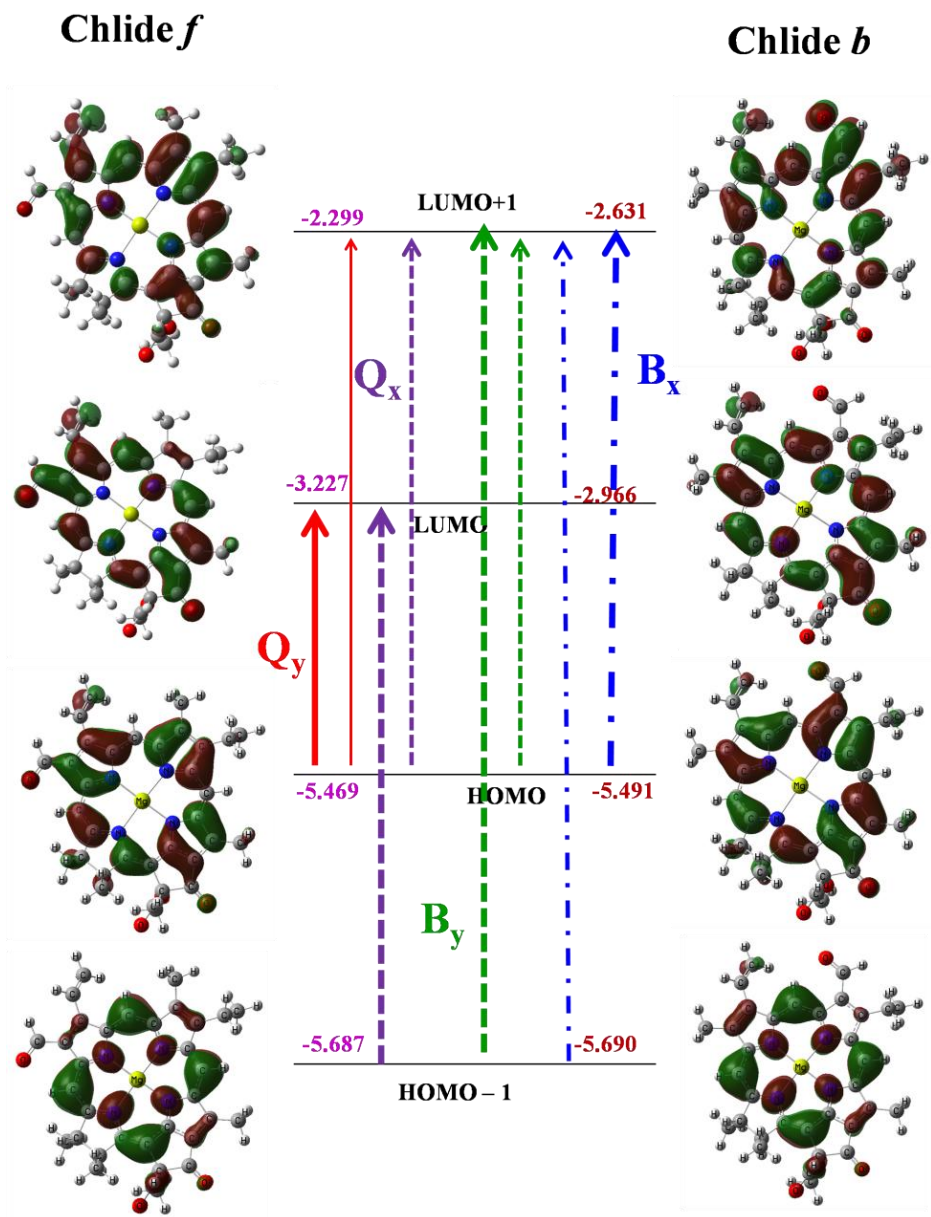

**Figure 5.** Orbital excitation plots of Chlide *f* and Chlide *b*. Corresponding MO pictures and MO energies (eV) are shown. The strength of the excitation line reflects its contribution for that absorbance band. Isocontour values used for molecular orbital plots is = 0.02 e (Bohr)$^{-3}$

In addition to the $Q_y$ band, the $Q_x$ bands were observed at 545 nm, 552 nm and 579 nm with lower intensity as shown in figure 3a, 3b for Chlides *a*, *b* and *f* respectively. The $Q_x$ band in all the Chlides has a major contribution from the HOMO-1 to LUMO transition. The red shift in $Q_x$ band in Chlide $f$ is due to the same reason as found in $Q_y$ band.

The changes in the – CHO local contribution in the specific molecular orbital is also shown in the higher energy $B_x$ (370-400 nm) and $B_y$ (300-350 nm) bands, which have the major contributions from the excitations HOMO to LUMO+1 and HOMO−1 to LUMO+1, respectively (see Table 4 and figure 5). Although HOMOs in both the Chlides are nearly equal in energy, the LUMO + 1 orbital is more stablilized in Chlide $b$ than in Chlide $f$ due to the extra contribution of –CHO in Chlide $b$. This extra stabilization of LUMO + 1 in Chlide $b$ causes a blue-shift in the $B_x$ (32 nm) and $B_y$ (23 nm) bands of Chlide $f$ compared to Chlide $b$.

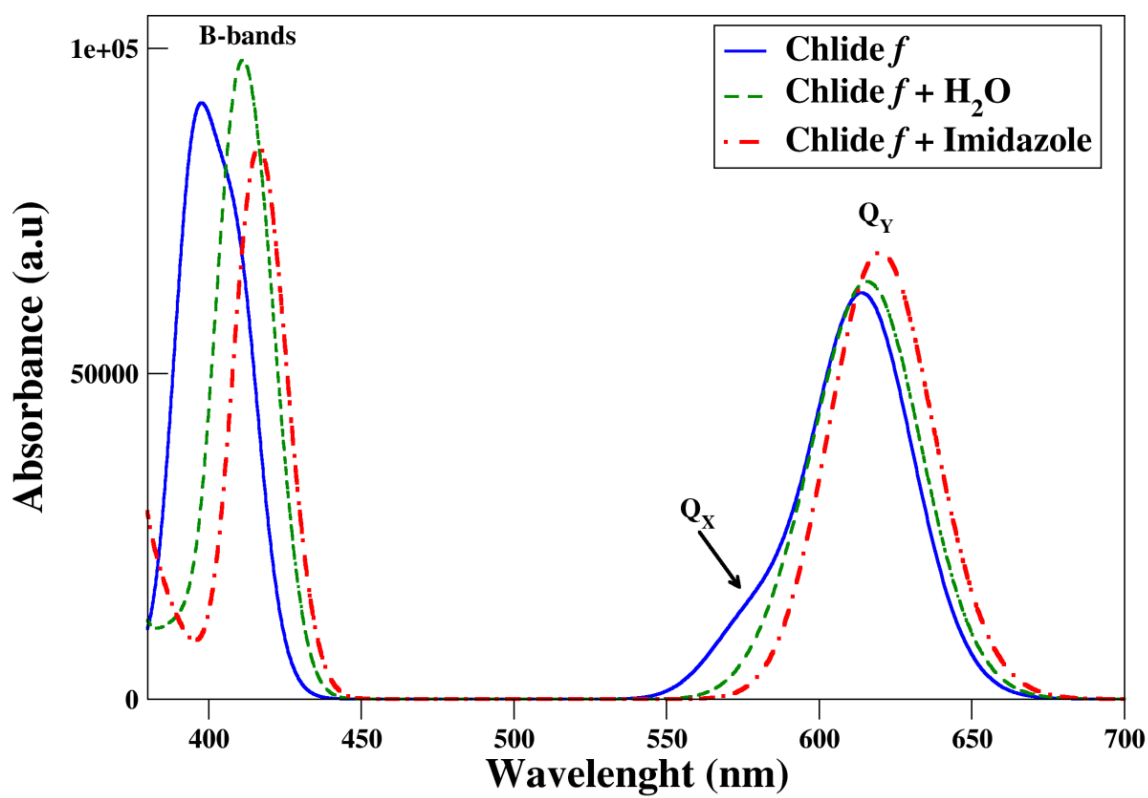

**Figure 6.** Computed absorption spectra of Chlide $f$ and $H_2O$, Imidazole ligated Chlide $f$.

Note that, all four excitation bands have major contribution from the π orbital of macrocycle chlorin ring, and are not affected by axial ligation (see figure S1). However, all bands of Chlide $f$ are red–shifted on axial ligation (see figure 6 and figure S2). Our calculations show a shift of about 5 − 10 nm in the $Q_y$ band and 9 – 30 nm in the $Q_x$ band depending on the nature of the fifth ligand, which can be understood, again, by observing the changes in the energies of HOMOs and LUMOs up on addition of axial ligand (Table 4). All the MOs of axial ligated Chlide *f* are destabilized during the addition of axial ligands, which is the reason for a red shift in $Q_x$, $Q_y$, $B_x$ and $B_y$ bands (see Table 4 and figure 4).

**Conclusions**

Density functional theory and time dependent density functional theory computations have been carried out to understand the different electronic excited state properties of Chlide *f*, compared to its structural isomer Chlide *b*. For Chlide *f* and Chlide *b*, computed $^1$H-NMR chemical shielding value clearly show a downfield chemical shift for formyl hydrogen than $H_a$, $H_b$ and $H_c$, which compares well with the experimental values. Axial ligation to Chlide *f* distorts Mg atom from its plane, however, it follows the similar trend of the chemical shielding values as in Chlide *f*, where the formyl hydrogen is always in downfield when compared to the $H_a$, $H_b$ and $H_c$.

In general, low energy $Q_x$, $Q_y$ and high energy $B_x$ and $B_y$ bands are observed for all chls, which have a major contribution from macrocycle ring localized π molecular orbitals (HOMO, HOMO-1, LUMO and LUMO+1). The computed $B_x$, $B_y$, $Q_x$ and $Q_y$ bands excitation energies of Chlide *a*, Chlide *b* and Chlide *f* are in good agreement with available experimental, and in certain cases, with earlier theoretically reported values. We have found that the LUMO of Chlide *f* is stabilized due to the larger contribution of –CHO substituent, which is the reason for its red shift of $Q_x$ and $Q_y$ bands when compared with Chlide *b*. Even in the presence of axial ligation, the macrocycle chlorin ring plays a major role in the excitation and the redox processes. Axial ligation shows red–shift in the all bands. Our results provide evidence that the Chlide *f*-$CH_3COOH$ with more positive reduction potential values can act as electron acceptor and Chlide *f*-Imidazole with less positive reduction potential values can act as electron donor, when compared with other possible ligands.

ACKNOWLEDGMENT

SKP acknowledges DST, government of India for the financial support.

SUPPORTING INFORMATION

The absorption spectra for axial ligated Chlide *f* and their corresponding orbital excitation plots are shown in the supporting information

**SUPPORTING INFORMATION**

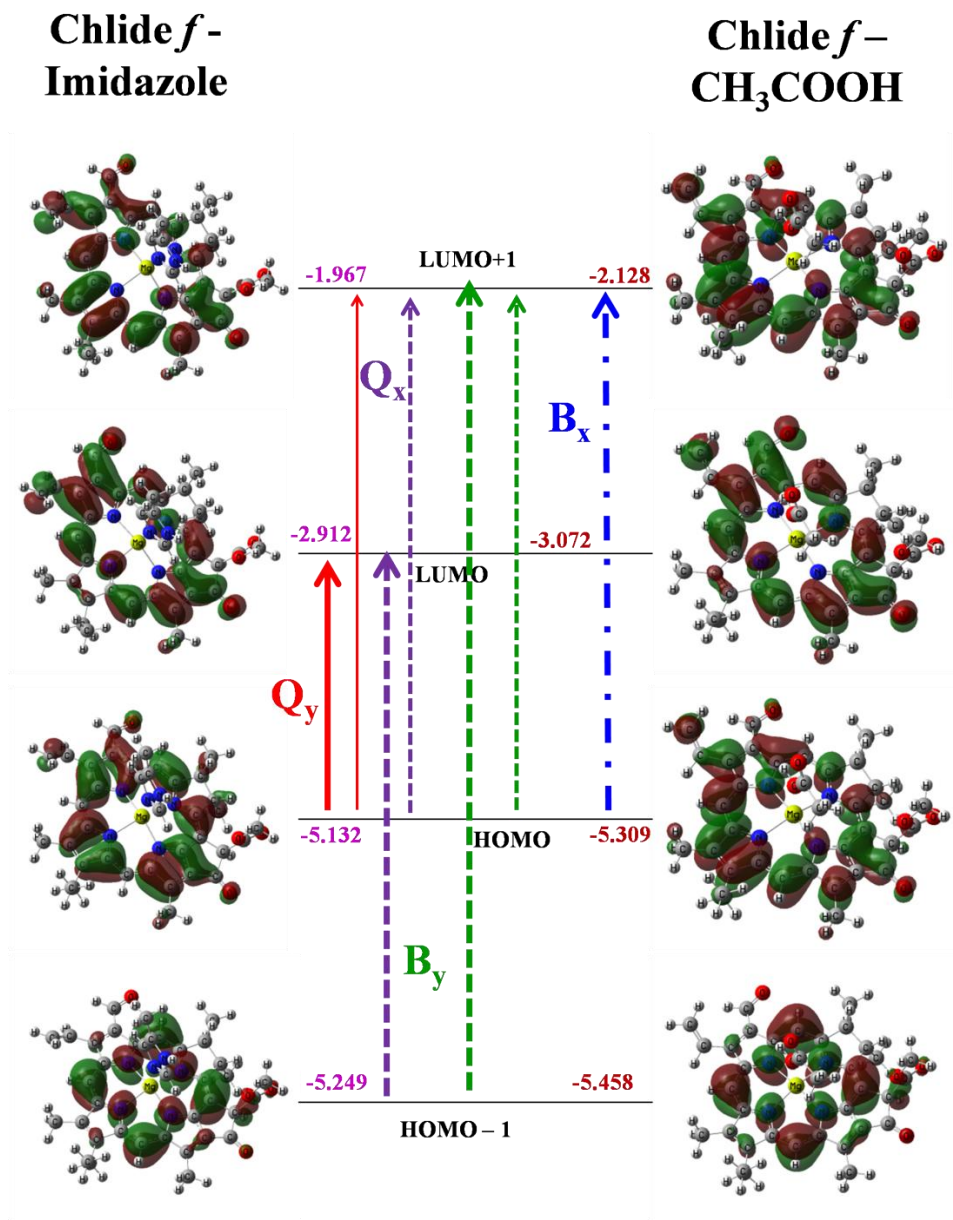

**Figure S1.** Orbital excitation plots of Chlide *f* -Imidazole and Chlide *f*-CH$_3$COOH. Corresponding MO pictures and MO energies (eV) are shown. The strength of the excitation line reflects its contribution for that absorbance band. Note that this plot clearly shows that the axial ligations do not have much contribution to absorption spectra. Isocontour values used for molecular orbital plots is = 0.02 e Bohr$^{-3}$

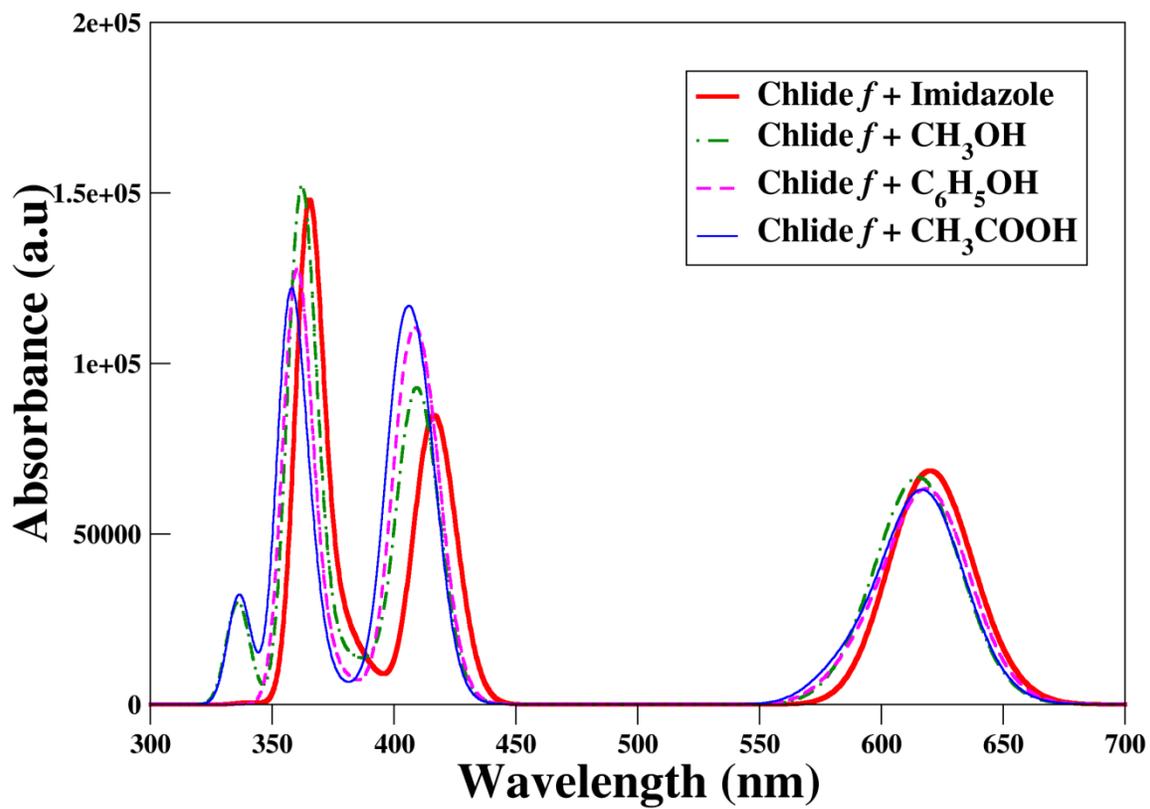

**Figure S2.** Absorption spectra of axial ligated Chlide *f*- (a) Imidazole, (b) CH$_3$OH, (c) C$_6$H$_5$OH and (d) CH$_3$COOH molecules. FWHM = 1000 cm$^{-1}$

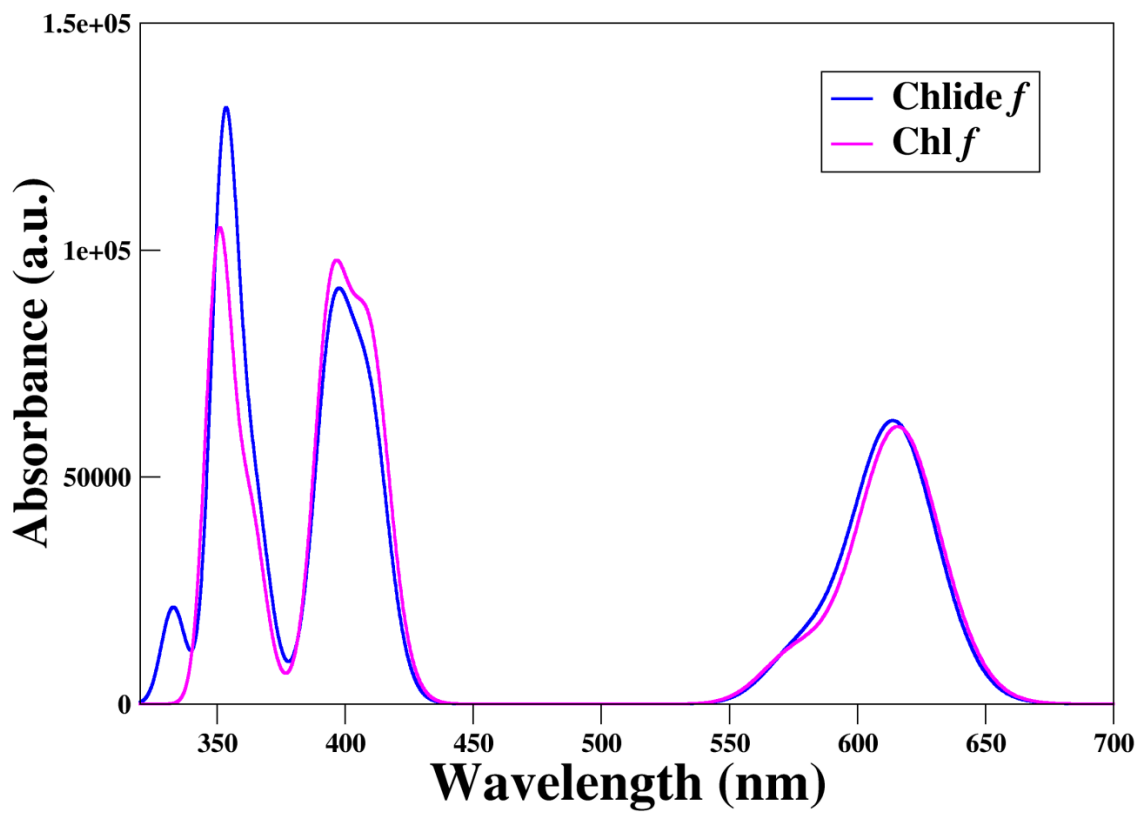

**Figure S3.** Absorption spectra of chl *f* and Chlide *f* molecules. FWHM = 1000 cm$^{-1}$

In order to see whether the consideration of the substituent at c17 of ring II will change the absorption spectrum of Chlide *f*, we performed calculations on chl *f* (i.e. with the substituent at the 17$^{th}$ carbon of ring II) with the same level of theory used for Chlide *f*. As expected, there is no much change in the position of the bands after the addition of c17 substituent to Chlide *f*. Line width in figure S3 is 1000 cm$^{-1}$.

The changes in the position of bands can be seen in the Figure S3, S4 and in Table S1. As can be noted from the Table S1and from figure S4, all the bands in chl *f* are red/blue shifted within 1-2 nm.

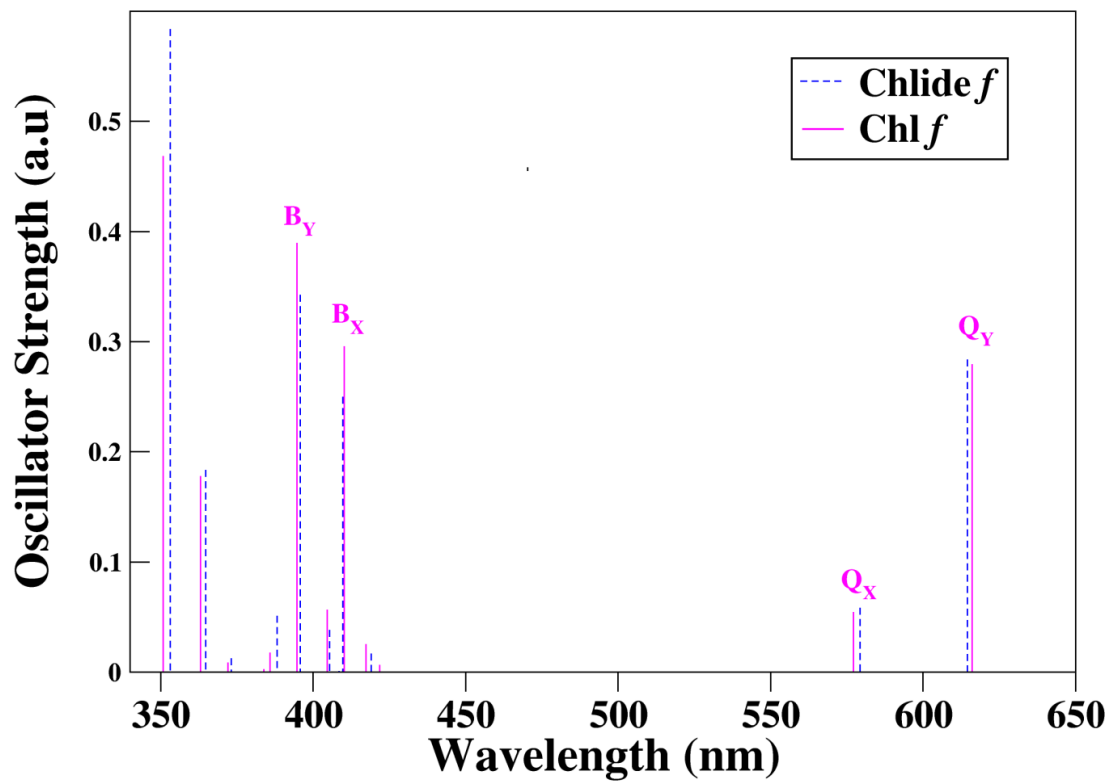

**Figure S4.** TDDFT computed electronic transitions of chl *f* and Chlide *f* molecules. Bands are assigned.

| Molecules | Nature | Excitation Wavelength (nm) | Major molecular orbital contributions |
|---|---|---|---|
| Chl *f* | $Q_y$ | 616 | H to L (0.72) |
|  | $Q_x$ | 577 | H-1 to L (0.72) and H to L + 1 (0.20) |
|  | $B_y$ | 395 | H−1 to L+1 (0.62) |
|  | $B_x$ | 351 | H to L+1 (0.50) |
| Chlide *f* | $Q_y$ | 615 | H to L (0.72) |
|  | $Q_x$ | 579 | H-1 to L (0.73) and H to L + 1 (0.19) |
|  | $B_y$ | 396 | H−1 to L+1 (0.61) |
|  | $B_x$ | 409 | H to L+1 (0.51) |

**Table S2:** Q and B bands of Chl *f* and Chlide *f*. Here 'H' stands for HOMO and 'L' stands for LUMO.

**Table of Contents**

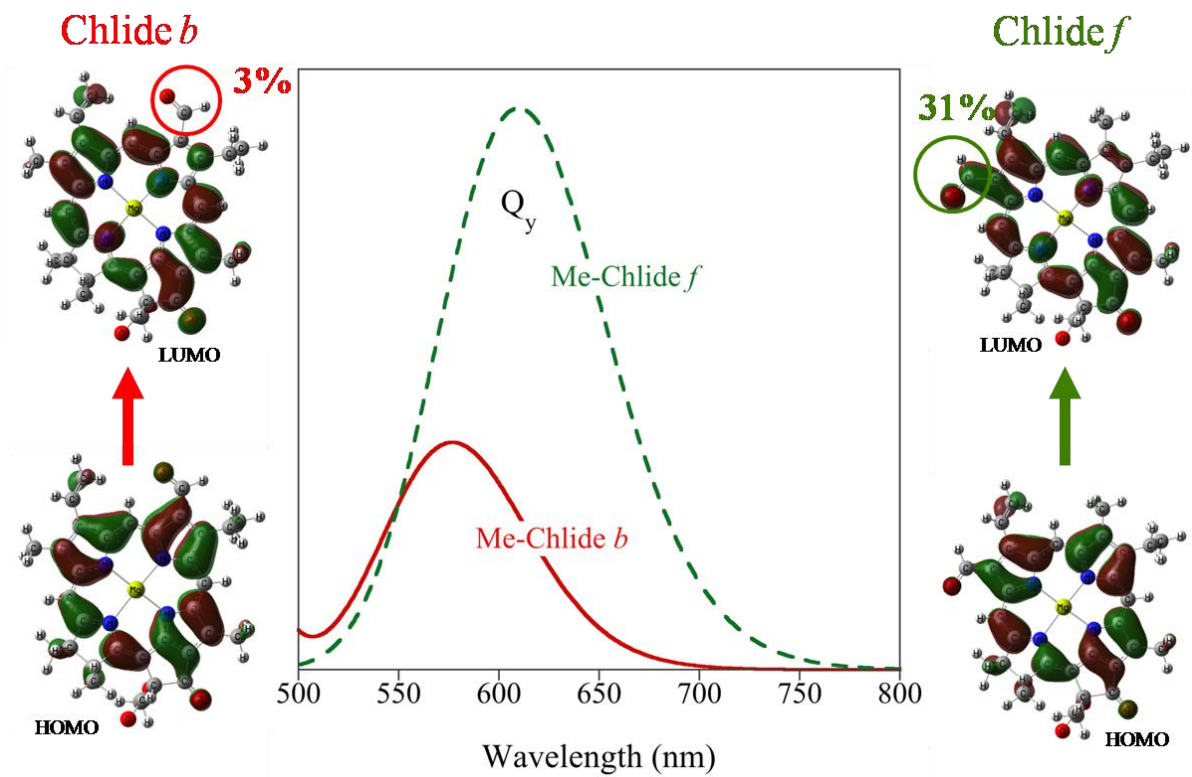

$Q_y$ absorption spectra of Chlide *f* and Chlide *b* with corresponding orbital excitations plots are shown. Note that –CHO contribution to LUMO of Chlide *f* and Chlide *b* are marked for clarification.